\documentclass{elsarticle}

\usepackage{graphicx}
\usepackage{amssymb}
\usepackage{a4wide}
\usepackage{amsmath,color}

\newcommand{\vph}{\ensuremath{\varphi}}
\newcommand{\vp}[1]{\ensuremath{\varphi_{#1}}}

\title{Cantilever Beam Equation for Almost Arbitrary Deflections: Derivation and Worked Examples}

\begin{document}

\begin{frontmatter}

\date{\today}

\title{Cantilever Beam Equation for Almost Arbitrary Deflections: Derivation and Worked Examples}

\author[focal]{Ale\v{s} Berkopec\corref{cor1}}
\ead{ales.berkopec@fe.uni-lj.si}

\cortext[cor1]{Principal Corresponding Author}
\address[focal]{FE, Tr\v{z}a\v{s}ka 25, Ljubljana, SI1000}

\begin{abstract}
We derived a non-linear 4$^\mathrm{th}$-order ordinary differential equation the solutions of which lead to
the exact shapes of the cantilever beam. The result of the equation in a non-dimensional form was found 
to depend on two parameters only: the angle of the beam at the fixed end, and the parameter encompassing
the material characteristics and geometry of the beam. The parameter space was explored in detail 
and the results were used to suggest the areas in which they could be applied.
\end{abstract}


\end{frontmatter}

\section{Introduction}

Cantilever is a homogeneous beam with one fixed end and one free end. In most practical applications 
it is attached horizontally to a vertical wall, allowed to bend under self-weight. 
Typical examples include steel beam as a part of a steel structure during 
construction of a high building, and the advancing part of a bridge while it is being built.~\cite{VazRodrigues20083024,Kwak2004767,Shooshtari2011454}
Lately, there have also been reports of its usage in nano-technology.~\cite{Arbat200976,Lee20071041,Chakraborty20061306} 

Theoretical results for a cantilever shape can be obtained analytically by solving the Euler-Bernoulli equation, 
or numerically by means of discretization of the beam domain, using finite elements method.~\cite{CurielSosa2009583} Both approaches have 
a drawback: the solutions of Euler-Bernoulli equation are valid only for small deflections, and the solutions with 
finite elements method require considerable resources in computational time, computational space, and programming skills.

In this article we present derivation of an ordinary non-linear differential equation of the fourth order, whose solution
describes the shape of the cantilever. We believe the procedure of obtaining the solution 
fits in between the two methods mentioned above. Firstly, the solution of the equation gives the stationary shape of a cantilever 
for (almost arbitrary) large deflections, and secondly, the solution is a result of a relatively simple, rather old, and widely 
accessible numerical algorithm.

In the first part we show how the governing fourth order differential equation is derived from second order Lagrangian for a one-dimensional 
structure with constant bending coefficient. The potential energy of the structure consists of a bending part, proportional to the curvature, 
and gravitational part. Minimization of the potential energy according to the calculus of variations leads to the
governing equation. We compare its solution with Euler-Bernoulli solution in terms of the height of the free end of horizontally fixed beam.
Because we introduce non-dimensional parameters, the shape of an arbitrary cantilever depends on two parameters only: its angle
at the fixed end, $\vph(0)$, and $e=\rho S g L^3/EI$. In the Results section we present two families of shapes in regard to the 
angle $\vph(0)$, and the geometrical properties of the free end as functions of both $e$ and $\vph(0)$. Possible applications of the 
solution can be found in Discussion.

\section{Theory}
\newcommand{\lp}{\ensuremath{\tilde{l}}} 

\begin{figure}[htb]
\centerline{\includegraphics{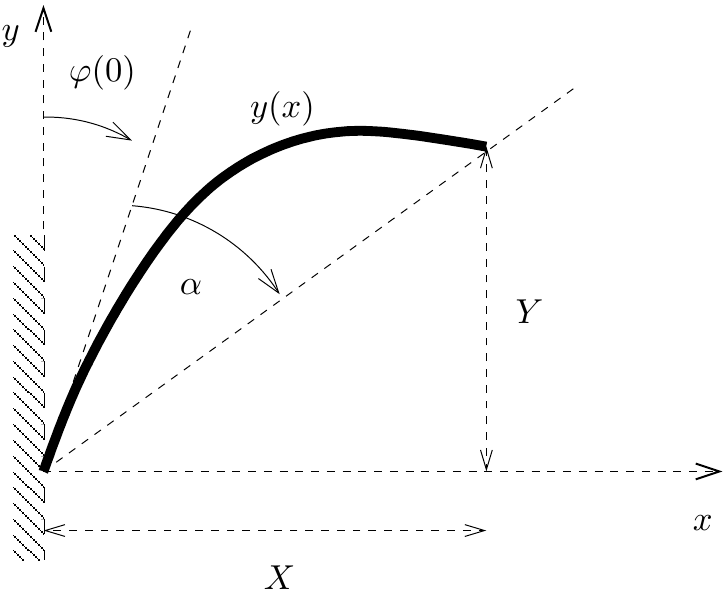}}
\caption{Geometry of the setup, where $y(x)$ denotes the shape 
of the cantilever beam.}
\label{fig:cbeam}
\end{figure}

A cantilever beam in this paper is a thin homogeneous beam of length $L$, 
fixed at one end, but otherwise free to bend under its self-weight.
The shape of the beam shall be given as a function $y(x)$, where $x$ is horizontal 
and $y$ is vertical component of the points on the beam (gravity points
in $-y$ direction). We choose the origin of the coordinate system 
$(x,y)$ at the fixed point of the beam, as shown in Fig.~\ref{fig:cbeam}.

The potential energy for a beam of a given shape $y(x)$ is a sum of a bending part and a gravitational part~\cite{Levan20051166}, or
\begin{equation}
W_\text{p}=\frac{\kappa}{2}\int_0^L C(\lp)^2\,\text{d}\lp+\rho_l g \int_0^L y(\lp)\,\text{d}\lp
\label{eq:contV}
\end{equation}
where $\kappa$ is a bending coefficient in units Jm, $\lp$ is an independent variable running along the beam from the fixed point 
at $\lp=0$ to the free end at $\lp=L$ in units of length, $C(\lp)=y^{\prime\prime}/(1+{y^{\prime}}^2)^{3/2}$ is a curvature 
of the beam at $\lp$ in units m$^{-1}$, $\rho_l$ is a mass per unit length in units kg/m, and $y(\lp)$ is a height of a beam 
point at $\lp$. 
In curvature $C(\lp)$ symbols $y^{\prime}$ and $y^{\prime\prime}$ denote first and second derivatives of $y(x)$ with respect 
to the independent variable $x$.

After we perform the substitutions $y\to Ly$, $x\to Lx$, $\text{d}\lp^2=L^2\text{d}t^2=\text{d}x^2+\text{d}y^2$, and $e=\rho_lgL^3/\kappa$,
the curvature changes into:
$$
C(\lp)^2=\frac{{y^{\prime\prime}}^2}{(1+{y^\prime}^2)^3}=\frac{1}{L^2}\cdot\frac{\ddot{y}^2}{1+\dot{y}^2}
$$
and the potential energy can be expressed in a more convenient form:
\begin{equation*}
W_\text{p}=\frac{\kappa}{2L}\int_0^1\bigg[\frac{\ddot{y}^2}{1-\dot{y}^2}+2\,e\,y\bigg]\,\text{d}t
\end{equation*}
The system reaches the stable equilibrium when potential energy is minimal, or when Lagrangian function
\begin{equation}
{\cal{L}}=\frac{\ddot{y}^2}{1-\dot{y}^2}+2\,e\,y
\label{eq:Lagrangian}
\end{equation}
solves the equation
\begin{equation}
-\frac{\partial^2}{\partial{t}^2}\bigg(\frac{\partial\cal{L}}{\partial\ddot{y}}\bigg)
+\frac{\partial}{\partial{t}}\bigg(\frac{\partial\cal{L}}{\partial\dot{y}}\bigg)
-\frac{\partial\cal{L}}{\partial{y}}=0
\label{eq:EulerLagrange}
\end{equation}
For $\cal{L}$ in Eq.~(\ref{eq:Lagrangian}) and the Euler-Lagrange Eq.~(\ref{eq:EulerLagrange}) we derive the governing
differential equation for $y(t)$ that reads:
\begin{equation}
\ddddot{y}=\frac{4\,(\dot{y}^2-1)\,\dot{y}\,\ddot{y}\,\dddot{y}-(3\dot{y}^2+1)\,\ddot{y}^3+(\dot{y}^2-1)^3e}{(\dot{y}^2-1)^2}
\label{eq:ddddoty}
\end{equation}
The range of the independent variable is $t\in[0,1]$. The non-dimensional 
co-ordinate $y$ is computed directly from 
Eq.~(\ref{eq:ddddoty}), while for $x$ we make use of 
$\dot{x}=\sqrt{1-\dot{y}^2}$.

In mechanics, it is common to use Young modulus $E$ and 
second moment of the area 
$I$, and not bending constant $\kappa$, like we did so far.
Conversion of constants, that leads to $E$ and $I$ in place of $e$, 
is performed by combining three equations:
the first one involves moment and curvature $C(\lp)=M(\lp)/(EI)$, 
the second and the third equation are equations for potential energy,
$\text{d}W_\text{p}=(1/2)\kappa\,C(\lp)^2\,\text{d}\lp$
and 
$\text{d}W_\text{p}=M(\lp)^2/(2EI)\,\text{d}\lp$.
It follows that the bending coefficient $\kappa$ equals a product of Young modulus $E$ and second moment of area $I$,
e.g.~$\kappa=EI$. The dimensionless parameter $e$ can now be expressed as
\begin{equation}
e=\frac{\rho_l\,g\,L^3}{EI}=
\frac{\rho\,S\,g\,L^3}{EI}
\label{eq:edef}
\end{equation}
where $\rho$ is mass density of the material, and $S$ is area of the cross section.

The Eq.~(\ref{eq:ddddoty}) reduces to the fourth order Euler-Bernoulli equation for small deflections
of horizontally bolted cantilever beam. A comparison of both solutions is presented in the next section.

\begin{figure}[hbt]
\centerline{\includegraphics[width=80mm]{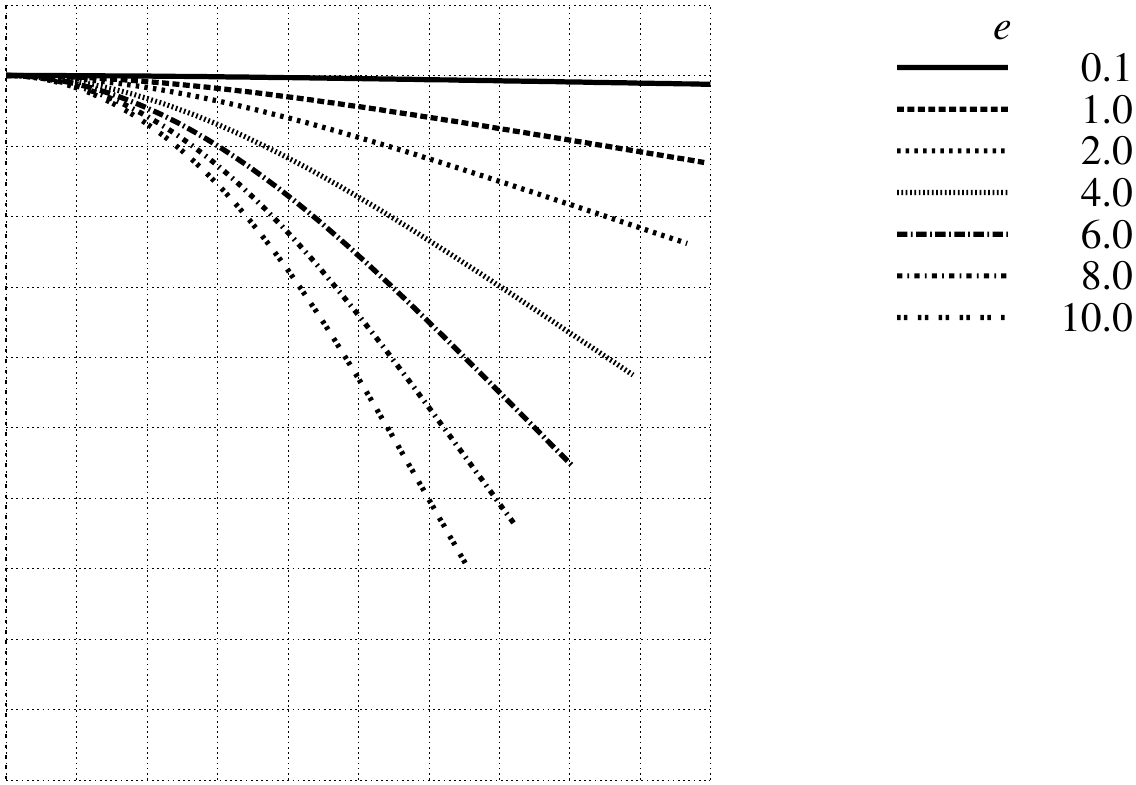}}
\caption{Stationary states of a cantilever beam bolted horizontally into a vertical wall.
For practical cases of the shapes of the beams that bend in the ways shown 
on this Figure consult Table \ref{tab:e}.
The shape of a cantilever that is bent most in this figure, that is for $e=10$, 
represents for example a shape of a 10~m long steel beam of circular 
cross section with diameter 24.55~mm, or a 1~m long circular hollow 
carbon fiber beam with 0.1~mm thickness and with outer 
diameter of 0.29~mm.}
\label{fig:cbeamhor}
\end{figure}

\section{Results}

The boundary conditions of the beam are $\dot{y}(0)=\cos\vph(0)$ at the fixed end, and
$\ddot{y}(1)=0$ and $\dddot{y}(1)=0$ at the free end. 
After we choose $y(1)$ and $\dot{y}(1)$, the integration of the 
differential equation Eq.~(\ref{eq:ddddoty}) is performed numerically 
backwards from $t=1$ to $t=0$. On completion, the resulting
function is shifted by $(x(0),y(0))$, so that the shape matches the 
geometry on Fig.~\ref{fig:cbeam} with $x(0)=y(0)=0$. 
One could compute the shape of the beam starting 
from $y(0)=0$ and $\dot{y}(0)$ and integrate Eq.~(\ref{eq:ddddoty}) 
while aiming for prescribed boundary conditions at $t=1$, 
but we presumed such procedure is less accurate and more time consuming, 
and was therefore not implemented. 
When $\dot{y}(0)$ is prescribed instead of $\dot{y}(1)$, we use bisection
algorithm with $\epsilon=10^{-12}$ to achieve the appropriate solution.

Integration of differential equation Eq.~(\ref{eq:ddddoty}) is done numerically.
Due to singularity at $\dot{y}=1$ the integrator, in our case {\slshape octave}'s
{\tt lsode} with stiff method~\cite{Acton,Recipes} and relative tolerance and absolute tolerance
$\sim1.5\cdot10^{-8}$, does not converge to a solution for small $\vph(0)$, for 
$|\cos\vph(0)|\gtrsim0.98$, or for severely bent cantilever $e\gtrsim10$ that occur 
for $|\cos\vph(1)|\gtrsim0.98$. The parameter space of interest here is therefore
$|\cos\vph(0)|\ge0.98$ and $e\le10$.

\begin{figure}[htb]
\centerline{\includegraphics[width=80mm]{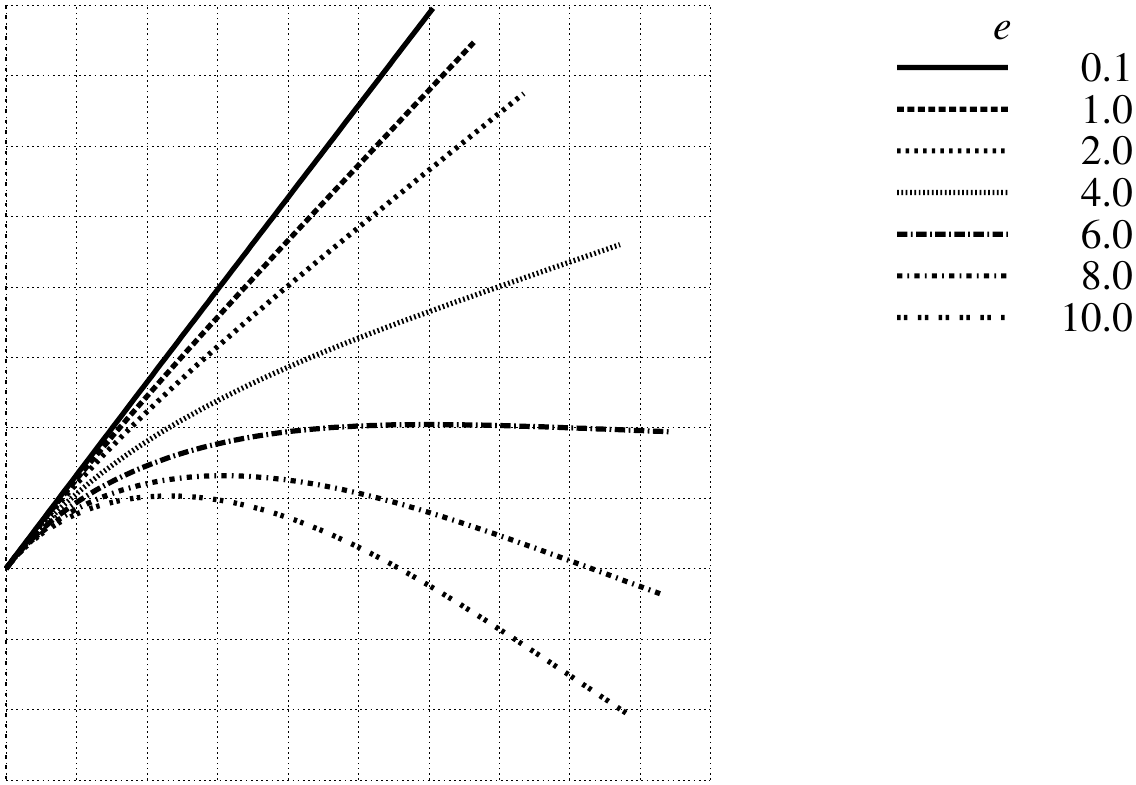}}
\caption{Stationary states of a cantilever beam bolted into a wall at $\cos\vph(0)=0.8$
($\vp0\approx36.87^\circ$, see Fig.~\ref{fig:cbeam}) for selected values of $e$.
For practical cases of the shapes of the beams that bend in the ways shown 
on this Figure consult Table \ref{tab:e}, see also Discussion for a suggestion of usage.
The shape of a cantilever that is bent most in this figure, that is for $e=10$, 
represents for example a shape of a 10~m long steel beam of circular 
cross section with diameter 24.55~mm, or a 1~m long circular hollow 
carbon fiber beam with 0.1~mm thickness and with outer 
diameter of 0.29~mm.}
\label{fig:cbeamtilt}
\end{figure}

In order to check our work with already known results we compare our solution with the solution of 
the Euler-Bernoulli (E-B) equation $EI\mathrm{d}^4\tilde{y}/\mathrm{d}\tilde{x}^4=q$, where $\tilde{y}$ is a deflection 
of the beam, and for a beam under influence of self-weight $q=-\rho_lg$.~\cite{Timosenko} Non-dimensional co-ordinates 
and parameters are introduced by transformations $\tilde{y}\to L\tilde{y}$ and $\tilde{x}\to L\tilde{x}$.
The equation then reads $\mathrm{d}^4\tilde{y}/\mathrm{d}\tilde{x}=-\rho S g L^3/EI=-e$
and its analytical solution is
$$
\tilde{y}=-\frac{e\tilde{x}}{4}\cdot\bigg[\frac{1}{6}\tilde{x}^3-\frac{2}{3}\tilde{x}^2+\tilde{x}-\frac{4c}{e}\bigg]
$$
where $c=\cos\vph(0)=\tilde{y}^\prime(0)$. Let $(\tilde{X},\tilde{Y})=(1,\tilde{y}(1))$ denote the co-ordinates of the cantilever free end 
obtained as a solution of the E-B equation, and let $(X,Y)=(x(1),y(1))$ denote the co-ordinates of the cantilever free end as a result
of Eq.~(\ref{eq:ddddoty}). The comparison of $\tilde{Y}$ and $Y$ for $c=0$ and a chosen set of parameters $e$ is shown in Table~\ref{tab:ebcomp}.

\begin{table}[hbt]
\centerline{
\begin{tabular}{|c||r|r|r|r|r|r|r|r|r|}
\hline
$e$         	
& $ 0.001$ & $ 0.01$ & $ 0.1$ & $ 1.0$ & $ 2.0$ & $ 4.0$ & $ 6.0$ \\
\hline
$\tilde{Y}$ 	
& $-0.0001$ & $-0.0013$ & $-0.0125$ & $-0.1250$ & $-0.2500$ & $-0.5000$ & $-0.7500$ \\
\hline
$Y$		
& $-0.0001$ & $-0.0012$ & $-0.0125$ & $-0.1235$ & $-0.2385$ & $-0.4252$ & $-0.5539$ \\
\hline
$|\tilde{Y}-Y|$
& $ 0.0000$ & $ 0.0000$ & $ 0.0000$ & $ 0.0015$ & $ 0.0115$ & $ 0.0748$ & $ 0.1961$ \\
\hline
\end{tabular}
}
\caption{Comparison of the heights of the free end for the solution of E-B equation and the solution of Eq.~(\ref{eq:ddddoty}) 
for $\cos\vph(0)=0$. For the images of the shapes see Fig.~\ref{fig:cbeamhor}.}
\label{tab:ebcomp}
\end{table}

\begin{table}[hbt]
\begin{tabular}{|r||r|r|r||r|r|r||r|r|r|}
 \multicolumn{10}{c}{\includegraphics{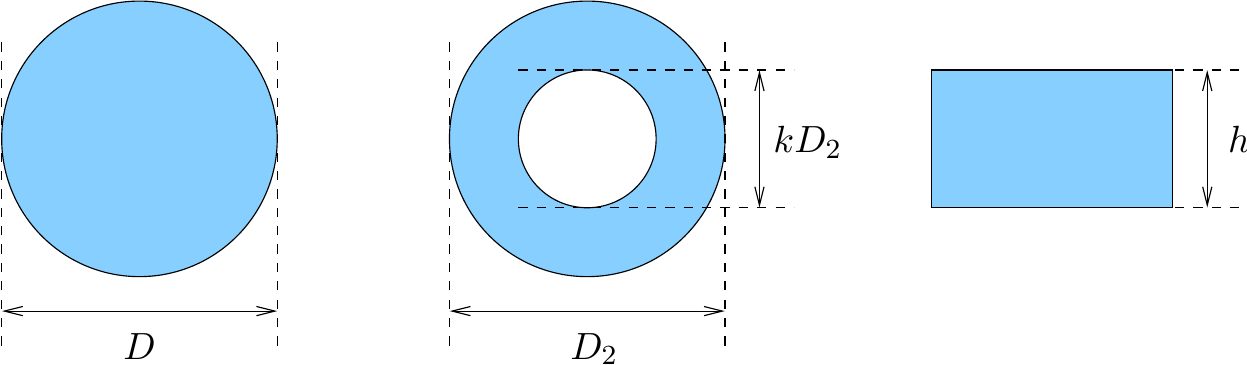}}\\
 &\multicolumn{9}{|c|}{\large$D={D_2}{\sqrt{1+k^2}}=\frac{2}{\sqrt{3}}h$ [mm]}\\ \cline{2-10}
         &\multicolumn{3}{|c||}{$L=  1$ m}&\multicolumn{3}{|c||}{$L= 10$ m}&\multicolumn{3}{|c|}{$L=100$ m}\\ $e$\ \ \  & {\slshape C--fiber}& {\slshape    glass}& {\slshape    steel}& {\slshape C--fiber}& {\slshape    glass}& {\slshape    steel}& {\slshape C--fiber}& {\slshape    glass}& {\slshape    steel}\\ \hline\hline
    0.01 &     9.70&    22.23&    24.55&   306.88&   702.83&   776.35&  9704.43& 22225.56& 24550.49\\ \hline
    0.10 &     3.07&     7.03&     7.76&    97.04&   222.26&   245.50&  3068.81&  7028.34&  7763.55\\ \hline
    1.00 &     0.97&     2.22&     2.46&    30.69&    70.28&    77.64&   970.44&  2222.56&  2455.05\\ \hline
    2.00 &     0.69&     1.57&     1.74&    21.70&    49.70&    54.90&   686.21&  1571.58&  1735.98\\ \hline
    4.00 &     0.49&     1.11&     1.23&    15.34&    35.14&    38.82&   485.22&  1111.28&  1227.52\\ \hline
    6.00 &     0.40&     0.91&     1.00&    12.53&    28.69&    31.69&   396.18&   907.35&  1002.27\\ \hline
    8.00 &     0.34&     0.79&     0.87&    10.85&    24.85&    27.45&   343.10&   785.79&   867.99\\ \hline
   10.00 &     0.31&     0.70&     0.78&     9.70&    22.23&    24.55&   306.88&   702.83&   776.35\\ \hline
\end{tabular}
\caption{Characteristic dimensions for three typical cross sections
of cantilever: circular, cylindrical hollow, and rectangular.
These are given for eight values of parameter $e$, 
three lengths $L$, and three different
materials that are linear over large range of strains, where 
C--fiber is an abbreviation for a carbon fiber. The material parameters
used in the table are as follows:
for carbon fiber $\rho=1800$~kg/m$^3$ and $E=300$~GPa,
for glass $\rho=2203$~kg/m$^3$ and $E=70$~GPa, and
for steel $\rho=7680$~kg/m$^3$ and $E=200$~GPa.
}
\label{tab:e}
\end{table}

\newcommand{\hep}{60mm}
\newcommand{\hmm}{10mm}
\begin{figure}[phtb]
\centerline{
\begin{minipage}{.5\textwidth}
  \hfil\includegraphics[height=\hep]{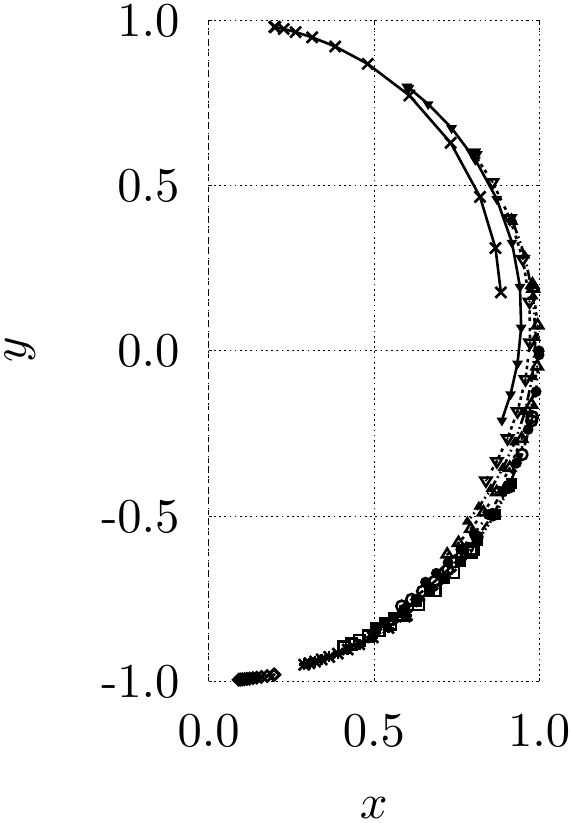}
\end{minipage}
\begin{minipage}{.5\textwidth}
\vspace{-10mm}
  \hfil\includegraphics[height=56mm]{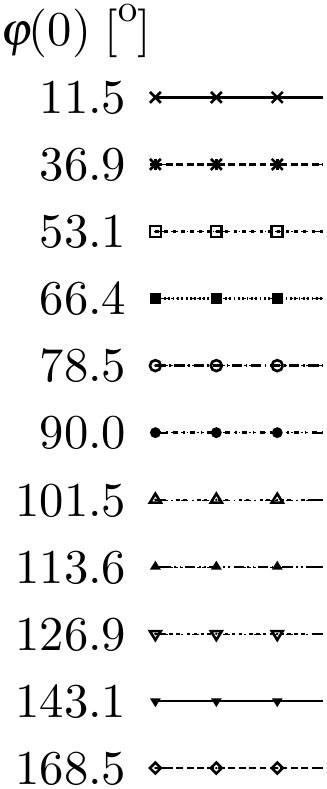}
\end{minipage}
}
\vskip5mm
\centerline{
\includegraphics[height=\hep]{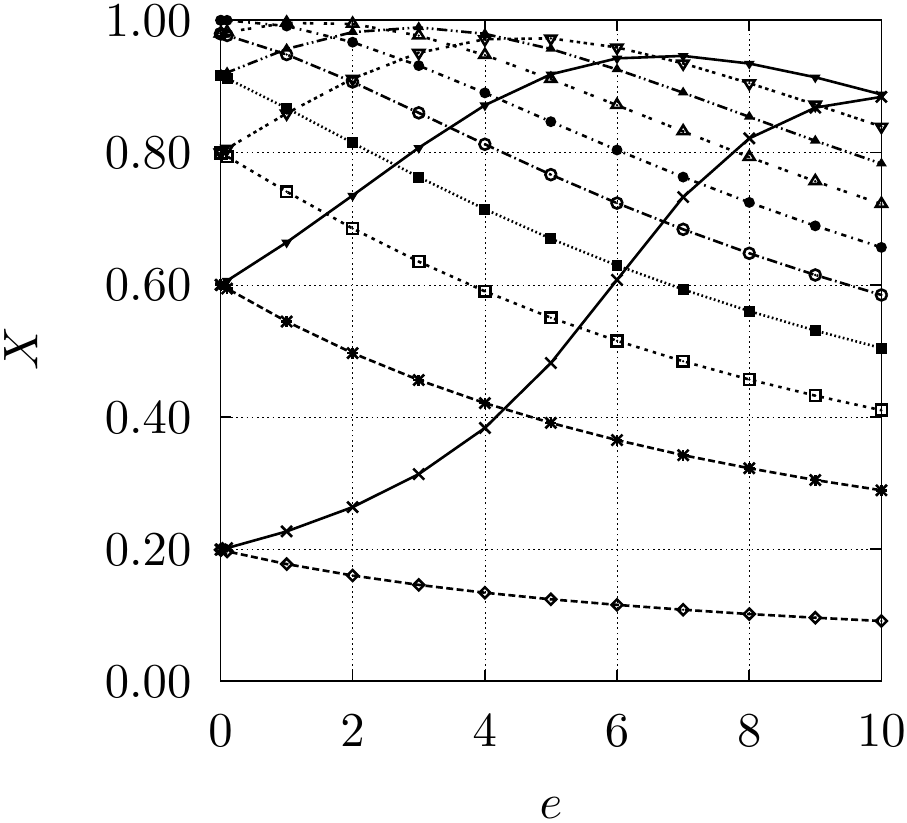}\hskip\hmm
\includegraphics[height=\hep]{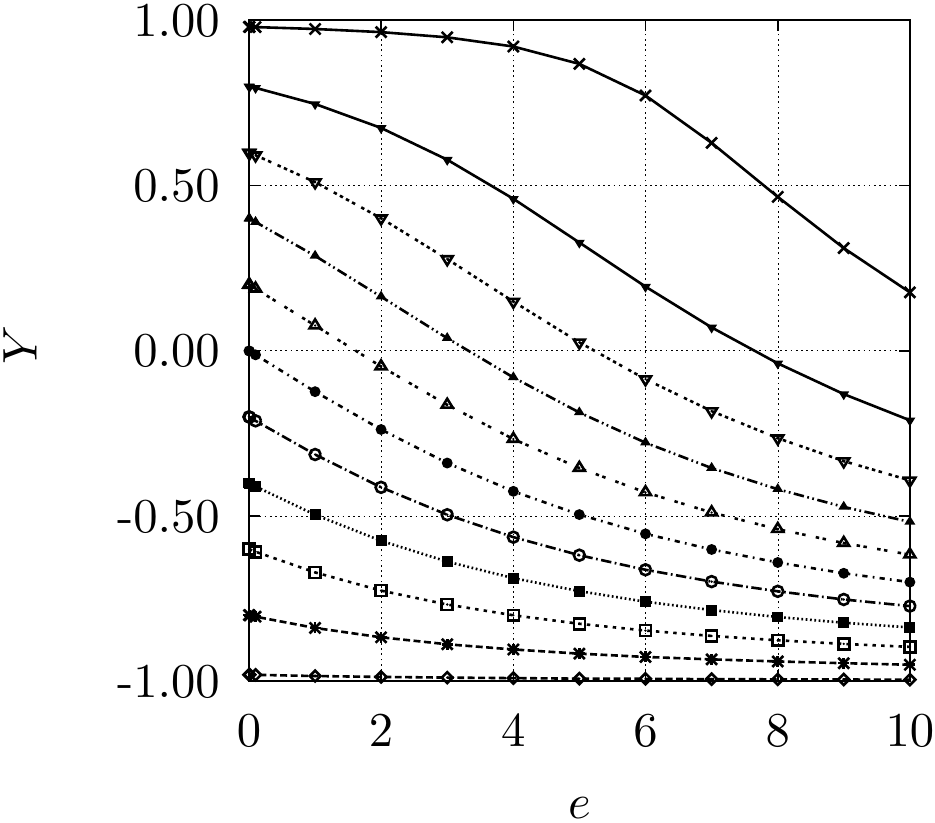}
}
\vskip5mm
\centerline{
\includegraphics[height=\hep]{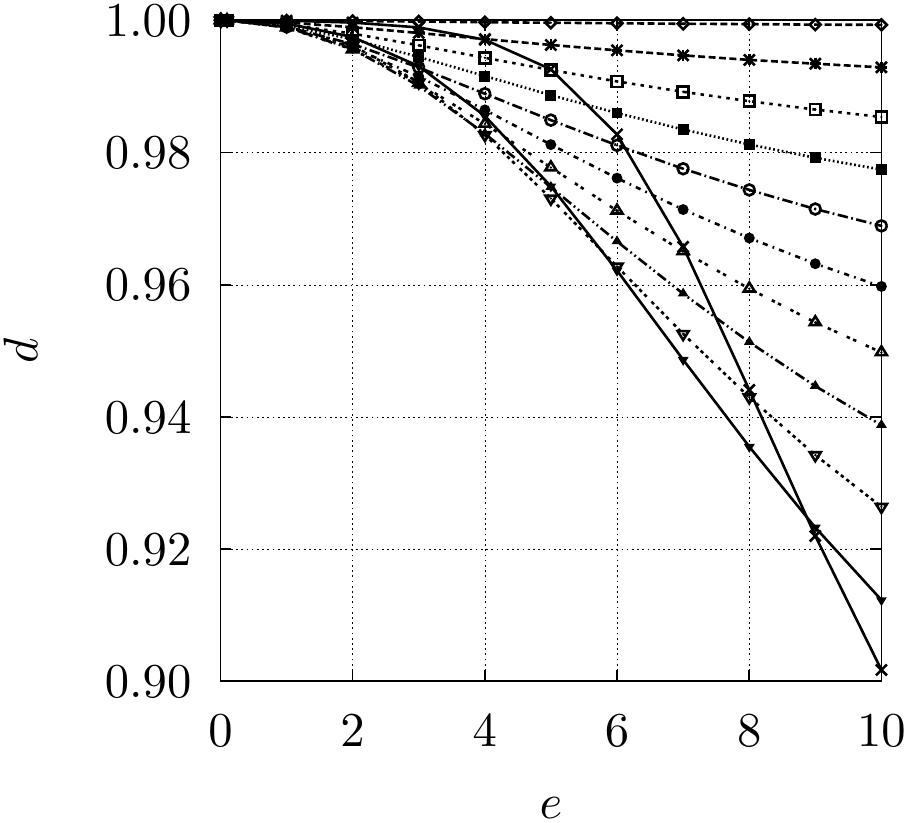}\hskip\hmm
\includegraphics[height=\hep]{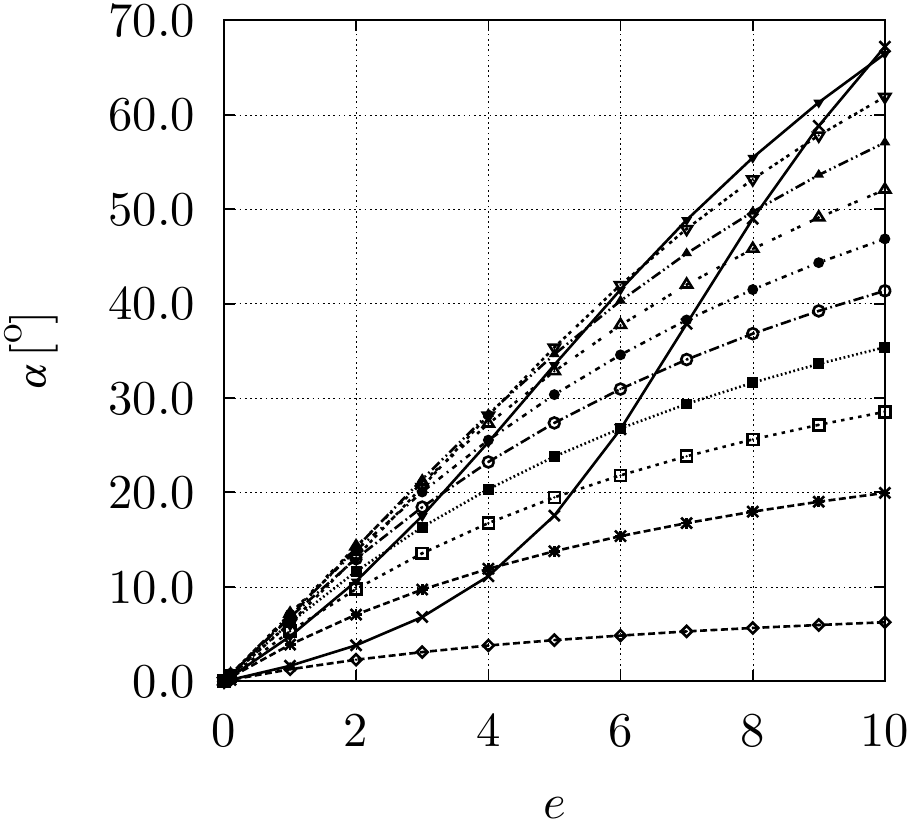}
}
\caption{The co-ordinates of the free end $(X,Y)$, distance from the origin $d=\sqrt{X^2+Y^2}$, and tilt angle of 
the free end $\alpha=90^\circ-\vph(0)-\arctan(Y/X)$. For graphical explanation see Fig.~\ref{fig:cbeam}. Top figures
show the co-ordinates of the free end $(X,Y)$ for different values of $\vph(0)$ and $e$ (left), and the legend
for $\vph(0)$ (right). The set of parameter values used to obtain these figures are $e\in[0.001,0.01,0.1,1,2,..,10]$ and 
$\cos\vph(0)=[-0.98,-0.8,-0.6,-0.4,..,0.8,0.98]$.}
\label{fig:xyda}
\end{figure}

\section{Discussion}

Besides the obvious application of Eq.~(\ref{eq:ddddoty}) where the shape of the cantilever beam is calculated from 
the geometry and the material parameters, e.g.~$(\rho,g,L,E,I)\to e\to(x,y)$, the exact solution of the
equation is also suitable to solve inverse problems. Since the shape of the beam is defined by $e$,
we may use the shape to get an estimation about any missing parameters in $e$, or 
$(x,y)\to(X,Y,\vp0)\to e\to(\rho,g,L,E,I)$. This discussion is dedicated mainly to two such examples.
The first elaborates an idea about a robust large-scale thermometer on the basis of temperature dependence of the
Young modulus $E$, and the second presents a procedure for determination of the cross section of a cantilever beam
in case when the cross section can not be determined otherwise.

The experimental evidence shows \cite{Tkalcec2004352,Kankanamge201126} the Young modulus $E$ is inversely proportional to the temperature.
In this example we consider a type of steel for which $E$ in the temperature range from $-200$~$^\circ$C to $800$~$^\circ$C
changes monotonously and approximately linearly from 210~GPa to 130~GPa. In Fig.~\ref{fig:xyda} we see that the height 
of the free end of the beam $Y$ is the most suitable candidate due to its largest and monotonous changes as a function of $e$. 
We make use of this fact at $\cos\vph(0)=0.8$, where the effect is emphasized between $e(Y=0.5)\approx4$ and $e(Y=0.0)\approx8$. 
Let us take a steel beam of circular cross section with diameter 1~mm and $E=200$~GPa. We get
$e=4$ when the length of the beam is $L=0.85$~m. Assuming the changes are linear, the rate of change in $Y$
as a function of temperature is $\Delta y\cdot L/\Delta T\approx 0.5\cdot 0.85/1000\approx0.4$~mm/K. Such a thermometer
is suitable for detecting changes in temperatures of order 100~K, and is usable also outside the temperature range 
stated above. The results of the shape and height of the free end $Y$ are sketched on Fig.~\ref{fig:xyda}. 
For such measuring system to be of practical use one should choose a cantilever beam whose material and geometrical properties have 
no hysteresis: the reversible changes in temperature should give reversible changes of $e$, and consequently reversible changes in the shape.

As another application of the cantilever beam equation consider a task of finding the inner diameter of a long flagpole of a circular shape.
In accordance with definition of the parameters in Table \ref{tab:e}, the parameter we seek in this task is $k$, a quotient of the inner 
diameter and the outer diameter. The parameter $k$ is in the range $[0,1]$: for a non-hollow flagpole $k=0$ and for a flagpole made 
of a very thin walls $k\to 1$.
Apart from the inner diameter, other geometrical and all the material properties of the flagpole are known. The flagpole is made of steel 
with $\rho=7680$~kg/m$^3$ and $E=200$~GPa, the outer diameter is $D=50$~mm, and the length of the pole is $L=12$~m. We assume the flagpole 
can be seen and/or photographed, so from the shape of the flagpole we find its initial angle is $\vp0\approx18^\circ$, measured from 
the vertical direction, and the tilt of the tip is estimated to be $\alpha\approx7^\circ$. From $\vp0$ and $\alpha$,
and by considering Fig.~\ref{fig:xyda} (lower right) we find $e(\vp0,\alpha)\approx 3$. 
For a circular hollow beam $I=\pi D^4(1-k^4)/64$ and $S=\pi D^2(1-k^2)/4$. These give $I/S=D^2(1+k^2)/16$, 
and from Eq.~(\ref{eq:edef}) we obtain $I/S=\rho g L^3/Ee$. Combined together, these equations give:
\begin{equation}
k=\sqrt{\frac{16\rho g L^3}{EeD^2}-1}\approx0.623
\label{eq:k}
\end{equation}
In estimation of the error we assume that all the quantities except angles were measured with an accuracy of 0.1~\% or better, so their
contribution to the overall error in $k$ is neglected. The errors in the measurements 
of the angles $\vp0$ and $\alpha$ are assumed to be at most $\pm2\%$. 
We have used the angles to obtain an estimation of $e$, so Fig.~\ref{fig:xyda} 
is used again to give an estimation of the error in $e$  
imated to at most $\pm5\%$, or $e=3\cdot(1\pm0.06)$. From Eq.~(\ref{eq:k}) we get
$$
\frac{\partial k}{\partial e}\Bigg\vert_{e=3}=-\frac{\lambda}{2e^2\sqrt{\lambda/e-1}}
$$
where $\lambda={16\rho g L^3}/{ED^2}$. In our case $\lambda\approx4.2$ and $e=3$, so $\Delta k\approx-0.371\cdot\Delta e$. The procedure gives
for the lower and the upper estimates of $k$ values $0.623\pm0.02226$, or $0.623\cdot(1\pm0.04)$. Our calculation suggests that 
the flagpole is a hollow cylinder with inner radius of $31.15$~mm.

As there are not many linear materials that give considerably deflected shapes, like those for $e\ge2$, the 
beam shape can be used to determine the material of a beam by finding best approximation for $\rho/E$ from $e$.
In theory, a cantilever beam could be used as a sensor of gravity, too. For a given beam the parameter $e$
on the Moon is six times smaller than $e$ on the Earth, so the change in the shape according 
to Fig.~\ref{fig:xyda} can be vast for $e\in[1,10]$. But we believe there are better ways of doing it.

Finally, we emphasize the cubic dependence of $e$ upon $L$. This means that any errors in $L$ have
triple impact on our computational results when compared to errors in $\rho$, $E$, or $I$. However,
because the shape of the cantilever has the highest sensitivity to the changes in the length of the beam,
it could help determine the length scale of the image of a cantilever we are looking at. Between 
the everyday construction business in scale of meters and nano-technology with scale of micrometers 
there are six orders of magnitude difference, so in terms of the parameter $e$ there are eighteen. 
The whole variety of the shapes of the cantilever from stiff to severely bent are achieved within 
four orders of magnitude of the parameter $e$.


\begin{thebibliography}{10}

\bibitem{VazRodrigues20083024}
R.~Vaz Rodrigues, M.~Fernández Ruiz, and A.~Muttoni.
\newblock Shear strength of r/c bridge cantilever slabs.
\newblock {\em Engineering Structures}, 30(11):3024 -- 3033, 2008.

\bibitem{Kwak2004767}
Hyo-Gyoung Kwak and Je-Kuk Son.
\newblock Span ratios in bridges constructed using a balanced cantilever
  method.
\newblock {\em Construction and Building Materials}, 18(10):767 -- 779, 2004.

\bibitem{Shooshtari2011454}
Alireza Shooshtari, Hamed Kalhori, and Amirhasan Masoodian.
\newblock Investigation for dimension effect on mechanical behavior of a
  metallic curved micro-cantilever beam.
\newblock {\em Measurement}, 44(2):454 -- 465, 2011.

\bibitem{Arbat200976}
A.~Arbat, E.~Edqvist, R.~Casanova, J.~Brufau, J.~Canals, J.~Samitier,
  S.~Johansson, and A.~Diéguez.
\newblock Design and validation of the control circuits for a micro-cantilever
  tool for a micro-robot.
\newblock {\em Sensors and Actuators A: Physical}, 153(1):76 -- 83, 2009.

\bibitem{Lee20071041}
Dong-Weon Lee and Il-Kwon Oh.
\newblock Micro/nano-heater integrated cantilevers for micro/nano-lithography
  applications.
\newblock {\em Microelectronic Engineering}, 84(5-8):1041 -- 1044, 2007.
\newblock Proceedings of the 32nd International Conference on Micro- and
  Nano-Engineering.

\bibitem{Chakraborty20061306}
Anirban Chakraborty and Cheng Luo.
\newblock Fabrication and application of metallic nano-cantilevers.
\newblock {\em Microelectronics Journal}, 37(11):1306 -- 1312, 2006.

\bibitem{CurielSosa2009583}
J.L.~Curiel Sosa and A.J. Gil.
\newblock Analysis of a continuum-based beam element in the framework of
  explicit-fem.
\newblock {\em Finite Elements in Analysis and Design}, 45(8-9):583 -- 591,
  2009.

\bibitem{Levan20051166}
A.~Le van and C.~Wielgosz.
\newblock Bending and buckling of inflatable beams: Some new theoretical
  results.
\newblock {\em Thin-Walled Structures}, 43(8):1166 -- 1187, 2005.

\bibitem{Acton}
F.~S. Acton.
\newblock {\em Numerical Methods that Work}.
\newblock Harper \& Row Publishers and New York, 1970.

\bibitem{Recipes}
W.~H. Press and et~al.
\newblock {\em Numerical Recipes in Pascal}.
\newblock Cambridge University Press and Cambridge, 1990.

\bibitem{Timosenko}
S.~Timoshenko.
\newblock {\em Strength of materials, part 2, advanced theory and problems}.
\newblock 1930.

\bibitem{Tkalcec2004352}
I.~Tkalcec, C.~Azcoïtia, S.~Crevoiserat, and D.~Mari.
\newblock Tempering effects on a martensitic high carbon steel.
\newblock {\em Materials Science and Engineering A}, 387-389:352 -- 356, 2004.
\newblock 13th International Conference on the Strength of Materials.

\bibitem{Kankanamge201126}
Nirosha~Dolamune Kankanamge and Mahen Mahendran.
\newblock Mechanical properties of cold-formed steels at elevated temperatures.
\newblock {\em Thin-Walled Structures}, 49(1):26 -- 44, 2011.

\end{thebibliography}
\end{document}